\documentclass[12pt]{article}
\usepackage{cite,epsfig,amssymb,amsmath,graphicx,color}
\newcommand{\be}{\begin{equation}}
\newcommand{\ee}{\end{equation}}
\newcommand{\bear}{\begin{eqnarray}}
\newcommand{\eear}{\end{eqnarray}}
\newcommand{\beal}{\begin{align}}
\newcommand{\eeal}{\end{align}}
\newcommand{\ba}{\begin{array}}
\newcommand{\ea}{\end{array}}
\newcommand{\nn}{\nonumber}

\usepackage[normalem]{ulem}

\begin{document}

\vspace{9mm}

\begin{center}
{{{\Large \bf Partially Supersymmetric ABJM Theory with Flux}
}\\[17mm]
Yoonbai Kim$^{1}$,~~O-Kab Kwon$^{1}$,~~D.~D. Tolla$^{1,2}$\\[3mm]
{\it $^{1}$Department of Physics,~BK21 Physics Research Division,
~Institute of Basic Science,\\
$^{2}$University College,\\
Sungkyunkwan University, Suwon 440-746, Korea}\\[2mm]
{\tt yoonbai@skku.edu,~okab@skku.edu,~ddtolla@skku.edu} }
\end{center}

\vspace{20mm}

\begin{abstract}
Starting with generic Wess-Zumino type coupling
to constant four-form and the dual seven-form field strengths
in the ABJM theory, we obtain
mass-deformed theories with ${\cal N}=2,4$
supersymmetries. These theories contain massless scalar fields and allow the
implementation of the Mukhi-Papageorgakis Higgsing procedure.
Using this procedure, we connect the Higgsed 
theories to three-dimensional mass-deformed SYM theories. These are
also connected by the four-dimensional ${\cal N}=1^*,2^*$
mass-deformed SYM theories through dimensional reduction.
We classify the three-dimensional mass-deformed SYM theories 
of ${\cal N}=1,2,4$ supersymmetry, of which
a few cases of ${\cal N}=1,2$ are connected neither by MP Higgsing procedure nor
dimensional reduction.
\end{abstract}

\newpage

\tableofcontents

\section{Introduction}
Various three-dimensional supersymmetric gauge theories have attracted
much interest as the theories describing the low energy dynamics
of multiple M2/D2-branes with and without background fluxes.
Much of recent interests are focused on the superconformal Chern-Simons matter
theory of Aharony-Bergman-Jafferis-Maldacena (ABJM)~\cite{Aharony:2008ug},
which describes the dynamics of M2-branes on
$\mathbb{C}^4/\mathbb{Z}_k$ orbifold singularity.
It was known that the circle compactification of this theory via the
Mukhi-Papageorgakis (MP) Higgsing procedure~\cite{Mukhi:2008ux}
leads to the three-dimensional ${\cal N}=8$ super
Yang-Mills (SYM) theory~\cite{Pang:2008hw,Kim:2011qv}, which is the low energy
effective theory of multiple D2-branes.
(See also Refs.~\cite{Mukhi:2011jp,Jeon:2012fn}.) Though the circle compactification
of the ${\cal N}=6$ supersymmetry-preserving mass-deformed ABJM (mABJM)
theory~\cite{Hosomichi:2008jb,Gomis:2008vc} can also be taken into account,
the MP Higgsing procedure cannot be implemented as a method of the circle compactification.
 This is because in the ${\cal N}=6$ mABJM thoery  all the scalar fields
are massive and the bosonic potential does not
involve any flat direction allowing an
infinitely large vacuum expectation value of the scalar fields. The latter is a crucial
requirement for the application of MP Higgsing procedure.

The origin of the mass-deformation in the ${\cal N}=6$ mABJM theory
is identified by the presence of Wess-Zumino (WZ) type coupling to special type of constant four-form
and the dual seven-form field strengths~\cite{Lambert:2009qw, KKNT}
in the infinite M2-brane tension limit.
As we discussed in the previous paragraph, one cannot apply the MP Higgsing
procedure to the ${\cal N}=6$ mABJM theory.
In this regard, we construct some supersymmetric mABJM
theories with flat directions, which let the MP Higgsing procedure possible.
Subsequently, we relate the resulting theories after the MP Higgsing to three-dimensional
mass-deformed super Yang-Mills (mSYM) theories.
To be specific, we start from the gauge-invariant WZ-type
coupling~\cite{KKNT,Allen:2011pm} in the ABJM theory and then apply the formalism
to a generic constant field strength in the infinite M2-brane tension limit.

By appropriate choices of the fluxes we
construct mABJM theories preserving ${\cal N}=2,4$ supersymmetries.
An intriguing aspect of the partially supersymmetric
mABJM theories is the fact that they always contain certain number
of massless scalar fields which result in some flat directions of the bosonic potential.
We show that the MP Higgsing of the ${\cal N}=2$ mABJM
theory leads to a mSYM theory, with the same number of supersymmetry.
This mass-deformed theory is equivalent to one
of the three distinct three-dimensional mSYM theories, which contain
one massless vector multiplet and three massive matter
multiplets~\cite{Kim:2011qv}.
The three distinct theories are
obtained by making different choices of the mass parameters of the
six massive fermionic fields of the matter multiplets. Similarly, we
show that the ${\cal N}=4$ mABJM theory is equivalent to a unique ${\cal N}=4$
mSYM in three dimensions. We also notice
that one of the three distinct ${\cal N}=2$ mSYM theories, but not
the one obtained from the ${\cal N}=2$ mABJM theory, is equivalent to
the one from the dimensional reduction of the four-dimensional ${\cal N}=1^*$
mSYM theory studied by Polchinski-Strassler~\cite{Polchinski:2000uf}.
The ${\cal N}=4$ mSYM is also equivalent to the one from the dimensional
reduction of the ${\cal N}=2^*$ mSYM theory in four-dimensions.
In the framework of gauge/gravity correspondence, three-dimensional mSYM
theories have been studied
in Refs.~\cite{Bena:2000zb,Bena:2000fz,Bena:2000va,Ahn:2001nw,Hyun:2003se}.

The remaining part of the paper is organized as follows. In section
\ref{ABJM-H} we study the deformation of the ABJM theory with
generic WZ-type couplings to constant background fluxes.
For later convenience we single out only the WZ-type coupling which survives
in the limit of infinite tension of M2-brane.
In section \ref{sec3} we
appropriately choose the fluxes in order to preserve certain amount
of supersymmetry. In section \ref{sec4} we apply the MP Higgsing
procedure to the partially supersymmetric mABJM theories and obtain the corresponding
mSYM theories. We then study the classification of these theories in
relation with the dimensional reductions of four-dimensional mSYM
theories. The detailed procedure of the dimensional reduction of the
${\cal N}=1^*$, $2^*$ theories is included in appendix A. Section
\ref{sec5} is devoted to discussions and future research directions.

\section{ABJM Theory with Constant Flux}\label{ABJM-H}

The ABJM action~\cite{Aharony:2008ug} is given by a Chern-Simons matter theory with ${\cal N}=6$ supersymmetry and U($N$)$\times$U($N$) gauge symmetry,
\begin{align}\label{ABJMact}
S=\int d^3x{\cal L}_{\rm ABJM}=\int d^3x\,\left({\cal L}_0 + {\cal L}_{{\rm CS}} + {\cal
L}_{{\rm ferm}} +{\cal L}_{{\rm bos}} \right),
\end{align}
where
\begin{align}
{\cal L}_0 &= {\rm tr}\left(-D_\mu Y_A^\dagger D^\mu Y^A +
i\Psi^{\dagger A} \gamma^\mu D_\mu \Psi_A\right),
\label{L0}\\
{\cal L}_{{\rm CS}} &= \frac{k}{4\pi}\,\epsilon^{\mu\nu\rho}\,{\rm
tr} \left(A_\mu \partial_\nu A_\rho +\frac{2i}{3}A_\mu A_\nu A_\rho
- \hat{A}_\mu \partial_\nu \hat{A}_\rho -\frac{2i}{3}\hat{A}_\mu
\hat{A}_\nu \hat{A}_\rho \right),
\label{LCS} \\
{\cal L}_{{\rm ferm}} &= -\frac{2\pi i}k{\rm tr}\Big( Y_A^\dagger
Y^A\Psi^{\dagger B}\Psi_B -Y^A Y_A^\dagger\Psi_B \Psi^{\dagger B}
+2Y^AY_B^\dagger\Psi_A\Psi^{\dagger B} -2Y_A^\dagger
Y^B\Psi^{\dagger A}\Psi_B
\nn \\
&\hskip 1.9cm  +\epsilon^{ABCD}Y^\dagger_A\Psi_BY^\dagger_C\Psi_D
-\epsilon_{ABCD}Y^A\Psi^{\dagger B}Y^C\Psi^{\dagger D} \Big),
\label{Lfe} \\
{\cal L}_{{\rm bos}} &=\frac{4\pi^2}{3k^2}{\rm tr}\Big(
Y^\dagger_AY^AY^\dagger_BY^BY^\dagger_CY^C
+Y^AY^\dagger_AY^BY^\dagger_BY^CY^\dagger_C
+4Y^\dagger_AY^BY^\dagger_CY^AY^\dagger_BY^C
\label{Lbo} \\
&\hskip 2cm -6Y^AY^\dagger_BY^BY^\dagger_AY^CY^\dagger_C \Big).\nn
\end{align}
The four complex
scalar fields $Y^A$ $(A=1,2,3,4)$ represent the eight directions $X^I$ $(I=1,\cdots,8)$
transverse to the M2-branes with
\begin{align}\label{YA}
Y^A=X^A+iX^{A+4}.
\end{align}
This action has ${\cal N}=6$ supersymmetry with the following
transformation rules
\begin{align}\label{susy}
&\delta Y^A=i\omega^{AB}\Psi_B,\quad \delta
Y^\dagger_A=i\Psi^{\dagger B}\omega_{AB},
\nonumber\\
&\delta\Psi_B=-\gamma^\mu\omega_{AB}D_\mu
Y^A+\frac{2\pi}k\omega_{BC}\big(Y^CY^\dagger_AY^A-Y^AY^\dagger_AY^C\big)
+\frac{4\pi}k\omega_{AC}Y^AY^\dagger_BY^C,
\nonumber\\
&\delta\Psi^{\dagger B}=D_\mu
Y^\dagger_A\omega^{AB}\gamma^\mu+\frac{2\pi}k\omega^{BC}\big(Y^\dagger_AY^AY^\dagger_C
-Y^\dagger_CY^AY^\dagger_A\big)
-\frac{4\pi}k\omega^{AC}Y^\dagger_AY^BY^\dagger_C,
\nonumber\\
&\delta A_\mu=-\frac{2\pi}k\big(Y^A\Psi^{\dagger
B}\gamma_\mu\omega_{AB}+\omega^{AB}\gamma_\mu\Psi_BY^\dagger_A\big),
\nonumber\\
&\delta \hat A_\mu=-\frac{2\pi}k\big(\Psi^{\dagger
B}\gamma_\mu\omega_{AB}Y^A+Y^\dagger_A\omega^{AB}\gamma_\mu\Psi_B\big),
\end{align}
where $\omega^{AB}=-\omega^{BA}=(\omega_{AB})^*=
\frac{1}{2}\,\epsilon^{ABCD}\omega_{CD}$.

Since the ABJM theory describes low energy dynamics of $N$ stacked M2-branes, it is intriguing to consider this theory in the background of a constant transverse four-form and the
dual seven-form field strengths. Interaction between the M2-branes and the background
three-form gauge fields is depicted by the WZ-type coupling.
In the presence of a constant transverse four-form field strength $F_4$,
the components of the corresponding three-form gauge field $C_3$ have the following
transverse scalar dependence:
\begin{align}\label{constC3}
&C_{\mu\nu\rho},~~ C_{\mu\nu A},~~ C_{\mu AB},~~ C_{\mu A \bar B}~~
{\rm and ~ their~complex~conjugate~(c.c.)~ are~
constants,}\nonumber\\
&C_{ABC},\,\, C_{AB\bar C}\,\, {\rm and~ their~c.c.~ are~linear~in~ transverse~scalars.}
\end{align}
Here we employed the index notations of~\cite{Kim:2011qv}, where the unbarred indices are contracted with bifundamental fields while the barred ones are contracted with anti-bifundamental fields.
We can set the constant components of $C_3$ in \eqref{constC3} to zero by using the gauge transformation of the three-form gauge field, $\delta C_3 = d\Lambda_2$.
 In addition, one cannot construct U($N)\times$U($N$) gauge-invariant WZ-type coupling with linear $C_{ABC}$~\cite{KKNT}. Therefore, the only gauge-invariant WZ-type coupling for this particular choice of the three-form gauge field is read from the equation (2.3) of Ref.~\cite{KKNT},
\begin{align}\label{massC3}
S_{C}^{(3)} = \lambda\int d^3x\, &\frac{1}{3!}
\epsilon^{\mu\nu\rho}{\rm tr}\Big[ C_{A\bar B C} D_\mu Y^A D_\rho Y_B^\dagger D_\nu Y^C
+ ({\rm c.c.}) \Big],
\end{align}
where $\lambda=2\pi  l_{{\rm P}}^{3/2}$ and $l_{{\rm P}}$ is the Planck length.

The dual seven-form field strength $F_7$ is expressed in terms of $F_4$ as
\begin{align}\label{F7}
F_7=\ast F_4+\frac12 C_3\wedge F_4.
\end{align}
%where `*' represents the Hodge duality operation.
According to the argument of the previous paragraph, in the presence of the constant transverse $F_4$, the $C_3\wedge F_4$ term in \eqref{F7}
is linear in the transverse scalar, while the $\ast F_4$ term is constant.
Keeping this in mind, we notice the following transverse scalar dependence
for the six-form gauge field $C_6$:
\begin{align}\label{constC6}
&C_{\mu ABCDE},~C_{\mu ABCD\bar E},\cdots,
C_{\mu\nu ABCD},~ C_{\mu\nu ABC\bar D},\cdots ~
{\rm  are~
constants,}\nonumber\\
&C_{\mu\nu\rho A B C},~ C_{\mu\nu\rho A B\bar C},\cdots\, {\rm are~linear~in~ transverse~scalars,}
\nn \\
&C_{ABCDEF},~C_{ABCDE\bar F},\cdots {\rm are~quadratic~in~ transverse~scalars.}
\end{align}
Setting the constant components of $C_6$ in \eqref{constC6} to zero using gauge degrees of freedom,
we read the gauge-invariant WZ-type coupling from the equation (2.8) of Ref.~\cite{KKNT},
\begin{align}\label{act6}
S_{C}^{(6)} = -\frac{\pi}{k\lambda}\int &d^3x\, \frac{1}{3!}
\epsilon^{\mu\nu\rho} \left\{{\rm tr}\right\}\Big[ C_{\mu\nu\rho A
B\bar C} \beta^{AB}_{~C}
+\lambda^3\big(C_{ABCD\bar E\bar F}D_\mu Y^A D_\nu Y^B D_\rho Y_E^\dagger
\beta^{CD}_{~F} \\
&+C_{ABC\bar D\bar E\bar F}D_\mu Y^A D_\nu Y_D^\dagger D_\rho
Y_E^\dagger \beta^{BC}_{~F}
+C_{AB\bar C\bar D\bar E\bar F}D_\mu Y_C^\dagger D_\nu
Y_D^\dagger D_\rho Y_E^\dagger \beta^{AB}_{~F}\big) + ({\rm
c}.{\rm c}.) \Big],\nn
\end{align}
where $\beta^{AB}_{~C}\equiv\frac12(Y^AY_C^\dagger Y^B - Y^BY_C^\dagger Y^A)$.

In this paper, we consider the infinite tension limit of the M2-brane
($\lambda\to 0$), which was also considered in Ref.~\cite{Lambert:2009qw},
in order to turn off the coupling to gravity modes.
In this limit, the three-form
coupling in \eqref{massC3} and all the six-form couplings in
\eqref{act6} except the first term can be neglected. Then it is enough to take into account the following WZ-type
coupling,
\begin{align}\label{massC6}
S_{\rm WZ} = -\frac{\pi}{\lambda k}\int d^3x\, &\frac{1}{3!}
\epsilon^{\mu\nu\rho} {\rm tr}\big[ C_{\mu\nu\rho A B\bar C}
\beta^{AB}_{~C} + C^{\dagger}_{\mu\nu\rho AB\bar C}
(\beta_{~C}^{AB})^\dagger \big].
\end{align}
The six-form gauge fields which are linear in the transverse scalars
are given by
\begin{align}\label{CC6}
C_{\mu\nu\rho A B\bar C} =- 2\lambda
\epsilon_{\mu\nu\rho} T_{AB\bar C\bar D}Y_D^\dagger,\qquad
C^\dagger_{\mu\nu\rho A B\bar C} =-2\lambda
\epsilon_{\mu\nu\rho} T_{CD\bar A\bar B}Y^D,
\end{align}
where the complex-valued constant parameters
$T_{AB\bar C\bar D}= (T_{CD\bar A\bar B})^*$
are antisymmetric in the last two barred indices as well as the first two
unbarred indices.
Therefore, the action in \eqref{massC6} is simplified as
\begin{align}\label{massC6a}
S_{\rm WZ}=\frac{4\pi}{k}\int d^3x\,{\rm tr}\big(T_{AB\bar C\bar
D}Y_C^\dagger Y^AY_D^\dagger Y^B \big).
\end{align}
As we will see later, the quartic flux term of the ${\cal N}=6$ mABJM theory can be expressed by the WZ-type coupling \eqref{massC6a}.
In addition, different choice of constant flux can be taken into account in M-theory. If
the masses of the fermionic and bosonic fields are appropriately
chosen, the supersymmetry is partially preserved.

\section{Supersymmetry-preserving Mass-deformations}
\label{sec3}

In this section we discuss possible mass deformations of the ABJM theory
in the presence of the constant flux term \eqref{massC6a},
which preserve some amount of supersymmetry.
We start by introducing general gauge-invariant mass terms for scalar and fermion fields
in addition to the quartic WZ-type coupling \eqref{massC6a},
\begin{align}
& {\cal L}^m_{\rm bos}= -{\rm tr}\big(M_A^{~B} Y^A Y^\dagger_B\big) \quad {\rm
with}\,\, M_A^{~B}= (M_B^{~A})^*,
\nn \\
& {\cal L}^m_{\rm ferm}= -i{\rm tr}\big(\mu_A^{~B}\Psi^{\dagger
A}\Psi_B\big)\quad {\rm with}\,\,\mu_A^{~B}=(\mu_B^{~A})^*,
\label{fma}
\end{align}
where $ M_{A}^{~B}$ and $ \mu_{A}^{~B}$ are constant mass matrices.
Then the total Lagrangian is written as
\begin{align}\label{totLag}
{\cal L}_{{\rm tot}} = {\cal L}_{\rm
ABJM}+{\cal L}_{\rm WZ}+{\cal L}^m _{\rm bos}+{\cal L}^m_{\rm ferm}.
\end{align}

The corresponding supersymmetry transformation rules in \eqref{susy} for the scalar and gauge
fields are unaffected by the mass-deformation while those for the fermionic fields are modified by
\begin{align}\label{susym}
\delta'\Psi_A=\mu_A^{~B}\omega_{BC}Y^C,\qquad \delta'\Psi^{\dagger
A}=\mu_{B}^{~A}\omega^{BC}Y^\dagger_C.
\end{align}

From the invariance of the total Lagrangian \eqref{totLag}
under the total supersymmetry transformation $\delta +\delta'$, we fix the
values of $T_{AB\bar C\bar D},~M_A^{~B}$, and $\mu_A^{~B}$ according to the number
of supersymmetry.
Since $\delta {\cal L}_{{\rm ABJM}}=\delta'{\cal L}_{\rm WZ}= \delta' {\cal L}^m _{\rm bos}=0$,
we need to verify only the following invariance,
\begin{align}\label{susyinv}
\delta'{\cal L}_{\rm
ABJM}+\delta({\cal L}_{\rm WZ}+{\cal L}^m _{\rm bos}+{\cal
L}^m_{\rm ferm})+ \delta' {\cal L}^m_{\rm ferm}=0
\end{align}
up to total derivative. Using the supersymmetry transformation rules in \eqref{susy} and \eqref{susym}, one can verify \eqref{susyinv} under
the conditions,
\begin{align}
&\mu_A^{~A}=0,
\label{0th} \\
&\mu_A^{~B}\mu_B^{~C}\omega_{CD}-M_D^{~B}\omega_{AB}=0,
\label{1st} \\
&\mu_A^{~B}\omega_{CD}-\mu_C^{~B}\omega_{AD}-\mu_A^{~E}\delta_C^{~B}\omega_{ED}
+\mu_C^{~E}\delta_A^{~B}\omega_{ED}-2T_{AC\bar B\bar
E}\,\omega_{ED}=0.\label{2nd}
\end{align}

In order to check the validity of this general setup,
we apply it to the well-known maximal
supersymmetry preserving case~\cite{Hosomichi:2008jb,Gomis:2008vc}.
In this case SU(4) R-symmetry of the ABJM theory is broken to SU(2)$\times$SU(2)$\times$U(1) due to the mass matrix $\mu_{A}^{~B} = {\rm diag}(m,m,-m,-m)$ with a mass parameter $m$.
Then we determine the bosonic mass matrix and the nonvanishing components of the constant four-form tensor from the conditions \eqref{1st} -- \eqref{2nd} as
\begin{align}\label{N=6choice}
M_A^{~B} = m^2\delta_A^{~B},
\quad
T_{12\bar 1\bar 2} = -m,\quad T_{34\bar 3\bar 4} = m.
\end{align}
This result exactly matches the known result for the case of maximally supersymmetric
mABJM theory~\cite{Hosomichi:2008jb,Gomis:2008vc} and the choice \eqref{N=6choice} is unique up to field redefinitions~\cite{Kim:2009ny}.

In the subsequent two subsections, we consider the cases with flat directions in bosonic potentials, where some of scalar fields and corresponding superpartners are massless. In those cases some of the supersymmetries are necessarily broken. The models with ${\cal N}=2$ and ${\cal N}=4$ supersymmetries are constructed.

\subsection{${\cal N}=2$}\label{N=2}

Let us consider a bosonic potential which is flat along
only one complex scalar field. By supersymmetry, the corresponding single complex fermion field should be massless while the other three fermion fields remains to be massive.
Without loss of generality, we choose the fermionic mass matrix of the three massive fermionic fields as
\begin{align}
\mu_A^{~B}={\rm diag}(0,m_2,m_3,m_4),
\label{2m}
\end{align}
where $m_A$'s $(A=2,3,4)$ are real
mass parameters.  Then we notice that $m_2+m_3+m_4=0$ due to the condition \eqref{0th}.
In order to satisfy the conditions in \eqref{1st} and \eqref{2nd}, we
should keep  only one complex component of $\omega_{AB}$ and its complex conjugate nonvanishing.
To be specific, we choose nonvanishing $\omega_{14}$  and then $\omega_{23}$ is also nonvanishing by the reality condition of $\omega_{AB}$.
Substitution of these into \eqref{1st} determine $M_A^{~B}$ as
\begin{align}\label{N=2bosmass}
M_A^{~B}={\rm diag}(m^2_4,m^2_3,m^2_2,0),
\end{align}
and then  nonvanishing components of
$T_{AB\bar C\bar D}$ are determined by \eqref{2nd} as
\begin{align}
T_{12\bar1\bar 2} &= -T_{34\bar3\bar4}=\frac{m_2}2, \quad
T_{13\bar1\bar3} =-T_{24\bar2\bar4} =\frac{m_3}2, \quad
T_{14\bar1\bar4} =-T_{23\bar2\bar 3}= \frac{m_4}2. \label{2T}
\end{align}
 One may also choose different nonvanishing components, for instance, $\omega_{12}$ or $\omega_{13}$, however, the results are equivalent to the aforementioned case of the nonvanishing $\omega_{14}$,
up to field redefinition.

\subsection{${\cal N}=4$}\label{N=4}

In order to obtain the mass-deformed theory with ${\cal N}=4$ supersymmetry,
we have to turn on
the mass term for two complex scalar fields. Then two complex fermionic fields become massive while the other two are massless.
This implements an appropriate choice for the corresponding fermionic
mass matrix
\begin{align}
\mu_A^{~B}={\rm diag}(0,0,m,-m).
\label{4m}
\end{align}
The conditions in \eqref{1st} --\eqref{2nd} are satisfied
only when we keep two nonvanishing complex supersymmetric parameters and their complex conjugates.
One possible choice is nonvanishing
$\omega_{13}$ and $\omega_{14}$ and then $\omega_{24}$ and $\omega_{23}$ are also
nonvanishing.
With this choice we read the bosonic mass matrix from \eqref{1st},
\begin{align}\label{N=4bosmass}
M_A^{~B}={\rm diag}(m^2,m^2,0,0),
\end{align}
and the following nonvanishing components of $T_{AB\bar C\bar D}$ from
\eqref{2nd}
\begin{align}
&T_{13\bar1\bar3} =-T_{14\bar1\bar4}=T_{23\bar2\bar
3}=-T_{24\bar2\bar4}= \frac{m}2. \label{4T}
\end{align}
Like the ${\cal N}=2$ case in subsection \ref{N=2}, this choice is unique up to
field redefinition.

In the original ABJM theory it was conjectured that the ${\cal N}=6$ supersymmetry
is enhanced to ${\cal N}=8$ at Chern-Simons levels $k=1,2$~\cite{Aharony:2008ug}.
The existence of such additional ${\cal N}=2$ supersymmetries was verified in terms of
the monopole operators~\cite{Gustavsson:2009pm,Kwon:2009ar,Bashkirov:2010kz,Samtleben:2010eu}.
For $k>2$, the supersymmetry enhancement is not possible due to orbifolding.
On the other hand, in order to implement the MP Higgsing procedure one has to move the M2-branes away from the orbifold singularity and this leads to an enhancement of the supersymmetry~\cite{Kim:2011qv}.
For instance, after the MP Higgsing procedure,
the ${\cal N}=6$ supersymmetry of the ABJM theory is
enhanced to the ${\cal N}=8$ supersymmetry of the three-dimensional SYM theory.
The latter theory flows to the supersymmetry enhanced ABJM theory
on flat transverse space ($k=1)$ at the IR fixed point~\cite{Kapustin:2010xq}.
However, as we shall show in the next section, there is no supersymmetry enhancement after the MP Higgsing procedure in the ${\cal N}=2,4$ mABJM theories. This implies the absence of the supersymmetry enhancement in the ${\cal N}=2,4$ mABJM theories, unlike the ${\cal N}=6$ mABJM theory.

\section{Classification}
\label{sec4}

The dimensional reduction of the ABJM theory with U($N)\times$U($N$)
gauge symmetry~\cite{Aharony:2008ug} via the MP Higgsing procedure
\cite{Mukhi:2008ux} leads to the three-dimensional ${\cal N}=8$
SYM theory with U($N$) gauge symmetry~\cite{Pang:2008hw,Kim:2011qv}.
In Ref.~\cite{Kim:2011qv} we have shown that the Higgsing of the ABJM theory deformed by WZ-type couplings of constant fluxes
results in effective theories of D2-branes in the background of constant RR fluxes.
By supersymmetric completion, for few choices of constant fluxes, we obtained ${\cal N}=2,4$ mSYM theories.
In the pervious section we have found the ${\cal N}=2,4$ mABJM theories.
Since these theories possess bosonic potentials
with flat direction, the MP Higgsing procedure can be carried out for these cases.
The resultant theories are compared with the aforementioned ${\cal N}=2,4$
mSYM theories as discussed in Ref.~\cite{Kim:2011qv}.
The ${\cal N}=1$ mSYM theory in Ref.~\cite{Kim:2011qv} cannot be obtained
from the MP Higgsing procedure of mABJM theory due to the following reason.
In order to apply this procedure the bosonic potential
is required to involve at least one massless complex scalar field. After the Higgsing, this complex field turns to one dynamical real massless scalar field and one would-be Goldstone boson.
In fact the ${\cal N}=1$ mSYM theory of Ref.~\cite{Kim:2011qv}
does not possess any massless scalar field.

\subsection{MP Higgsing of the ${\cal N}=2,4$ mABJM Theories}

To pursue the MP Higgsing procedure~\cite{Mukhi:2008ux} we proceed by introducing vacuum
expectation value $v$ for the massless scalar $Y^4$ along a transverse direction
\begin{align}\label{vev}
Y^A=\frac v2 T^0\delta^{A4} +\tilde X^{A}+i\tilde X^{A+4},
\end{align}
where $\tilde X^{I}$'s $(I=1,2,\cdots,8)$ are Hermitian scalar fields.
Correspondingly  we  introduce Hermitian fermionic fields
$\tilde\psi_r~(r=1,2,\cdots,8)$ as
\begin{align}\label{PsiA}
\Psi_A= \tilde\psi_A+i\tilde\psi_{A+4}.
\end{align}
  When the vacuum expectation value $v$ is turned on, in the MP Higgsing procedure, the U($N)\times{\rm U}(N)$ gauge symmetry
is broken to U($N$) and the Hermitian scalar and fermionic fields transform in adjoint representation
of the unbroken U($N$).  Then taking double scaling limit
of the large $v$ and large Chern-Simons level $k$ with finite $v/k$,
the Yang-Mills coupling $g$ is identified as $g=2\pi v/k$ and
the matter fields are rescaled as $\tilde \phi \to \tilde\phi/g$ for dimensional reason.
The detailed procedures are explained in Ref.~\cite{Kim:2011qv}.

Application of the Higgsing procedure to the total Lagrangian \eqref{totLag} results in
\begin{align}\label{totYM}
\tilde{\cal L}^{{\cal N}=2,4}_{{\rm YM}} = \tilde{\cal L}^{{\cal N}=8}_{{\rm YM}}
+\frac{1}{g^2}{\rm tr}\left( i\tilde T_{ijk} \tilde X^i [\tilde X^j,\,\tilde X^k] - \tilde M_{ij}\tilde X^i\tilde X^j -i\tilde\mu_{rs}\tilde \psi_r\tilde\psi_s\right),
\end{align}
where $i,j,k =1,2,\cdots, 7,$ and
\begin{align}\label{N=8SYM}
\tilde{\cal L}_{\rm YM}^{{\cal N}=8}
&=\frac 1{g^2}{\rm tr}\Big(-\frac12 \tilde F_{\mu\nu}\tilde F^{\mu\nu}
-\tilde D_\mu \tilde X^i \tilde D^\mu \tilde X^i
+ i\tilde\psi_r\gamma^\mu \tilde D_\mu\tilde\psi_r
+ \frac12[\tilde X^i,\,\tilde X^j]^2
-\Gamma_{i}^{rs}\tilde\psi_r [\tilde X^{i},\tilde\psi_s]\Big).
\end{align}
The cubic interaction term in \eqref{totYM} is the result of the MP Higgsing
of the WZ-type coupling \eqref{massC6a} and the antisymmetric tensors $\tilde T_{ijk}$ are related to the constant four-form tensor $T_{AB\bar C\bar D}$ in \eqref{CC6} as follows:
\begin{align}\label{Uijk}
&\tilde T_{ab4} = \tilde T_{4a+4b+4}= \frac{2i}{3} \big(T_{a4\bar b
\bar4} -T_{b4\bar a \bar4}\big),\quad \tilde T_{a4b+4} = -\frac{2}{3}
\big(T_{a4\bar b \bar4} +T_{b4\bar a \bar4}\big),
\nn \\
&\tilde T_{abc} = -i \big(T_{ab\bar c\bar 4} - T_{c4\bar a\bar b}\big),\quad
\tilde T_{a+4b+4c+4} =T_{ab\bar c\bar 4}+T_{c4\bar a\bar b},
\nn \\
&\tilde T_{abc+4} =\frac{1}{3}\big(2 T_{a4\bar b\bar c} - T_{c4\bar
a\bar b}-2 T_{ab\bar c\bar 4} +T_{bc\bar a\bar 4} \big),
\nn \\
&\tilde T_{ab+4c+4} =\frac{i}{3}\big(2 T_{c4\bar
a\bar b} -T_{a4\bar b\bar c} -2 T_{ab\bar c\bar 4} +T_{bc\bar a\bar 4} \big),
\quad (a,b,c=1,2,3).
\end{align}
The bosonic and fermionic mass terms in \eqref{totYM} are obtained from the MP Higgsing of
the mass terms \eqref{fma}.

For the ${\cal N}=2$ theory of subsection \ref{N=2} we read the nonvanishing components of
$\tilde T_{ijk}$ as well as the fermionic and bosonic mass matrices from \eqref{2m}--\eqref{2T},
\begin{align}
&\tilde\mu_{rs}={\rm diag}(0,m_2,m_3,m_4,0,m_2,m_3,m_4),
\quad \tilde M_{ij}={\rm diag}(m_4^2,m_3^2,m_2^2,0,m_4^2,m_3^2,m_2^2),
\nn \\
&\tilde T_{145} = \frac23m_4,\quad \tilde T_{246} = -\frac23 m_3,\quad
\tilde T_{347} = -\frac23 m_2\quad  {\rm with}\,\, m_2+m_3+m_4=0.
\end{align}
Similarly for the ${\cal N}=4$ theory of subsection \ref{N=4}, we have those quantities from \eqref{4m}--\eqref{4T},
\begin{align}\label{N=4para}
 &\tilde\mu_{rs}={\rm diag}(0,0,m,-m,0,0,m,-m),\quad \tilde M_{ij}={\rm diag}(m^2,m^2,0,0,m^2,m^2,0),
\nn \\
&\tilde T_{145} = -\frac23 m,\quad \tilde T_{246} = -\frac23 m.
\end{align}

These results can be compared with the corresponding mSYM theories in Ref.~\cite{Kim:2011qv} with suitable field redefinitions and parameter choices.
More precisely, the ${\cal N}=2$ mSYM theory we obtained in this paper is equivalent to
that of Ref.~\cite{Kim:2011qv} if we make the following field redefinitions and
identifications of the mass parameters of the two theories:
\begin{align}\label{para37}
&\tilde X^1\to\tilde X^1,\quad\tilde X^2\to\tilde X^3,\quad\tilde X^3\to\tilde X^6,\quad\tilde X^4\to\tilde X^7,\quad\tilde X^5\to\tilde X^2,\quad\tilde X^6\to\tilde X^4,\quad\tilde X^7\to\tilde X^5,
\nn \\
&m_2\to \mu_3=\mu_4,\quad m_3\to \mu_5=\mu_6,\quad m_4\to \mu_7=\mu_8,
\quad {\rm with}\,\,\mu_3+\mu_5+\mu_7=0,
\end{align}
where $\mu_i$'s are the mass parameters used in Ref.~\cite{Kim:2011qv}.
We call this theory `${\rm D}=3\,\, {\cal N}=2$ mSYM I'.
Actually, in the case of Ref.~\cite{Kim:2011qv} one can make other two more choices
of mass parameters satisfying all the constraints imposed by supersymmetry.
We call these theories `${\rm D}=3\,\, {\cal N}=2$ mSYM II \& III'.
However, these choices cannot be related with the Higgsing of the ${\cal N}=2$ mABJM theory
by field redefinition.
The reason is the fact that in the ${\cal N}=2$ mABJM theory we have only three mass parameters
of the three massive complex fields while in the case of Ref.~\cite{Kim:2011qv}
we have six mass parameters of the six massive real fields.
Therefore, in the latter case there are more freedoms in choosing the mass parameters.
See the next subsection for the details.

The comparison for the ${\cal N}=4$ theories obtained here and
Ref.~\cite{Kim:2011qv} can be made by setting
$m_2=0\to\mu_3=\mu_4=0$ and using the same field redefinitions and
parameter choices as in \eqref{para37}. In this case other possible
choices of mass parameters in Ref.~\cite{Kim:2011qv} are also
identical to the choice in \eqref{N=4para} up to field
redefinitions. The dimensional reduction of the four dimensional
${\cal N}=2^*$ mSYM theory~\cite{Polchinski:2000uf}
also gives this ${\cal N}=4$ mSYM theory. (For the details
see Appendix \ref{N=1star}.) It is also important to recall that we started with a general setting of mass deformation in ABJM theory to obtain the ${\cal N}=4$ mABJM theory.
In these regards, the ${\cal N}=4$ mABJM
and mSYM theories discussed in this paper seem unique.

\subsection{Classification of the ${\cal N}=2$ mSYM theories}

In Ref.~\cite{Kim:2011qv} we have shown that introduction of the mass
deformation to the ${\cal N}=8$ SYM theory preserves ${\cal
N}=1,2,4$ supersymmetries depending on the choices of the fermionic
and bosonic mass parameters and components of the
antisymmetric tensor $\tilde T_{ijk}$. In this subsection we fully
classify ${\cal N}=2$ mSYM theories according to mass
parameter choices.

The ${\cal N}=1$ mSYM theory contains one massless
gauge boson and seven massive scalar fields. Together with their
superpartners which are one massless and seven massive fermionic
fields, these sets of fields form one ${\cal N}=1$ vector multiplet
and seven massive matter multiplets. Since all the massive scalar fields
belong to different multiplets, they are allowed to have different
masses of which the parameters are unrestricted unlike the higher
supersymmetry cases. In this reason, there is no candidate in the mABJM theory
to be linked with this ${\cal N}=1$ mSYM theory.

In the case of the ${\cal N}=2$ mSYM theory, the supersymmetry
invariance of the action requires
\begin{align}
&\tilde\mu_{rs}={\rm diag}(0,0,\mu_3,\mu_4,\mu_5,\mu_6,\mu_7,\mu_8),
\quad  \tilde M_{ij}={\rm diag}(\mu_8^2,~\mu_7^2,~ \mu_6^2,~
\mu_5^2, ~\mu_4^2,~\mu_3^2,0),
\nn \\
&\tilde T_{145}=\frac 13(\mu_3+\mu_6+\mu_7),\quad \tilde
T_{246}=\frac 13(\mu_3+\mu_5+\mu_7),\quad
\tilde T_{347}=\frac 13(\mu_5+\mu_6),\nonumber\\
&\tilde T_{127}=-\frac 13(\mu_7+\mu_8),\quad \tilde T_{136}=-\frac
13(\mu_4+\mu_5+\mu_7),\quad
\tilde T_{235}=-\frac 13(\mu_3+\mu_5+\mu_8),\nonumber\\
&\tilde T_{567}=-\frac 13(\mu_3+\mu_4),
\end{align}
where the mass parameters are constrained as,
\begin{align}\label{N=2cons}
&\mu_4^2=\mu_3^2,\quad \mu_6^2=
\mu_5^2,\quad\mu_8^2=\mu_7^2,
\quad\mu_3+\mu_4+\mu_5+\mu_6+\mu_7+\mu_8=0.
\end{align}
%%%%%%%%%%%%%%%%%%%%%%%%%%%%%%%%%%%%%%%%%%%%%%%%%%%%%%%%%%%%%%%
\begin{figure}%[!h]
\centerline{\epsfig{figure=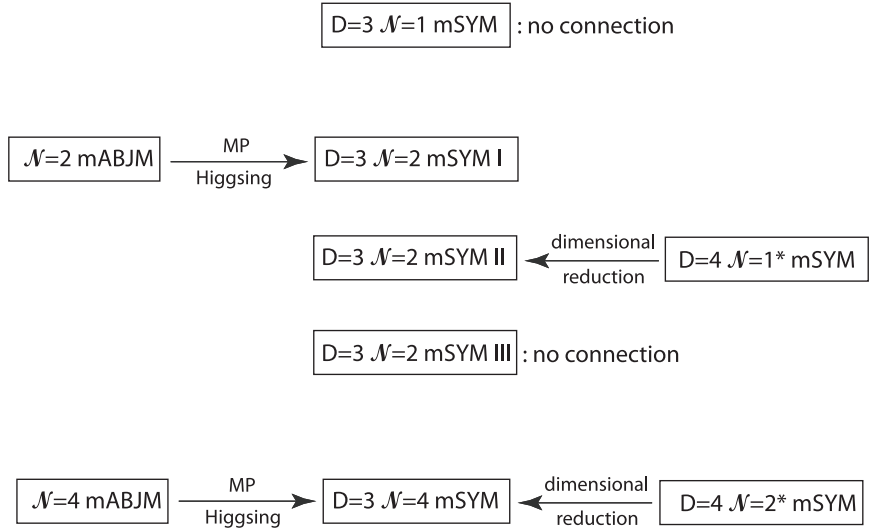,height=70mm}}
\caption{
\small Classification of ${\cal N}=1,2,4$ supersymmetric mass-deformed gauge theories in three dimensions and their relationship to the ${\cal} N=1^*,2^*$ mSYM in four-dimensions.}
\label{Fig1}
\end{figure}
%%%%%%%%%%%%%%%%%%%%%%%%%%%%%%%%%%%%%%%%%%%%%%%%%%%%%%%%%%%%%%%
There are three independent mass parameter choices satisfying the constraints
in \eqref{N=2cons},
\begin{align}\label{class}
\underline{{\rm case~I:}}\quad\quad
&\mu_3 = \mu_4,\quad \mu_5 = \mu_6,\quad \mu_7 = \mu_8
\quad {\rm with}~\mu_3 + \mu_5 + \mu_7=0,
\nn \\
&\tilde T_{145}=\tilde T_{246}=\tilde T_{136}=\tilde T_{235}=0,\quad
\tilde T_{347}=\frac 23\mu_5,\quad
\tilde T_{127}=-\frac 23 \mu_7,\quad
\tilde T_{567}=-\frac 23 \mu_3,
\nn \\ &
\nn \\
\underline{{\rm case~II:}}\quad\quad
&\mu_4 = -\mu_3,\quad \mu_6 = -\mu_5,\quad \mu_8 = -\mu_7,
\nn \\
&\tilde T_{145}=\frac 13(\mu_3-\mu_5+\mu_7),\quad
\tilde T_{246}=\frac 13(\mu_3+\mu_5+\mu_7),\quad
\tilde T_{136}=\frac 13(\mu_3-\mu_5-\mu_7),\quad
\nn \\
&\tilde T_{235}=\frac 13(-\mu_3-\mu_5+\mu_7),\quad
\tilde T_{347} = \tilde T_{127} = \tilde T_{567}=0,
 \nn\\ &
\nn \\
\underline{{\rm case~III:}}\quad\quad
&\mu_4 = -\mu_3,\quad \mu_6 =-\mu_8 = -\mu_7=\mu_5,
\nn \\
&\tilde T_{145}=\tilde T_{246}=\tilde T_{136}=-\tilde T_{235}=
\frac13\mu_3,\quad
\tilde T_{347}=\tilde T_{127}= \frac23\mu_5, \quad \tilde T_{567}=0.
\end{align}
As discussed previously the case I is identical to the Higgsed ${\cal N}=2$
mABJM theory of the previous subsection through the field redefinitions in \eqref{para37}.
As shown in appendix A, the case II is obtained as a result of the dimensional
reduction of the four-dimensional ${\cal N}=1^*$ mSYM theory~\cite{Polchinski:2000uf}.
The case III is a ${\cal N}=2$ mSYM theory
which can be connected to neither ${\cal N}=2$ mABJM theory through the MP Higgsing nor the ${\cal N}=1^*$ mSYM theory through
dimensional reduction. The reason why the cases II and III are not related with
the Higgsed ${\cal N}=2$ mABJM theory of subsection \ref{N=2}
is the following.
For the latter case, any set of two fermionic fields belonging to the same supermultiplet
is inherited from the real and imaginary components of a complex fermionic field
in the original mABJM theory. Therefore, they have the same masses.
For the former cases,
the fermionic fields in the same multiplet have either the same or opposite signs
for their mass parameters as indicated in \eqref{class}.
For convenience we summarize classification of the mSYM theories
in the diagram of Fig. 1.

\section{Conclusion}\label{sec5}

In this paper we classified some parity-preserving
three-dimensional supersymmetric mass-deformed gauge theories.
In the ABJM theory, we introduced a generic WZ-type coupling to
constant four-form and dual seven-form field strengths
in the limit of infinite M2-brane tension. We showed, with appropriate choice
of the fermionic and bosonic mass terms, such deformed ABJM theory
possesses ${\cal N}=2,4,6$ supersymmetries.
In Ref.~\cite{Kim:2011qv} we already constructed three distinct mSYM theories in three dimensions,
which are one ${\cal
N}=1$, three ${\cal N}=2$, and one ${\cal N}=4$ mSYM
theories.
Here we verified that one of the three ${\cal N}=2$ theories
and the ${\cal N}=4$ theory are obtained through the MP Higgsing of the ${\cal N}=2$ and
${\cal N}=4$ mABJM theories, respectively.
One of the remaining ${\cal N}=2$ theories is obtained by dimensional reduction of the
four-dimensional ${\cal N}=1^*$ theory,
while the ${\cal N}=4$ mSYM theory is also
obtained by the dimensional reduction to the ${\cal N}=2^*$ theory.
The third ${\cal N}=2$ and the ${\cal N}=1$ mSYM
theories are not connected by the MP Higgsing of the mABJM theory or the dimensional reduction of
the four-dimensional mSYM theory.

We may extend our analysis in this paper to the cases of the parity-violating
three-dimensional gauge theories, such as the ${\cal N}=3$ level-deformed ABJM theory developed by Gaiotto and Tomasiello (GT)~\cite{Gaiotto:2009mv} (see also Ref.~\cite{Fujita:2009kw}).
As the ABJM theory does, the GT theory allows the supersymmetry-preserving
mass-deformation~\cite{Kwon:2011nv} and the circle compactification via the MP Higgsing procedure~\cite{Go:2011bs}.
Utilizing these properties, one can construct
the less supersymmetric mass-deformed GT theories with flat directions which implement
the MP Higgsing procedure. This analysis may shed some light on
M-theory brane configuration of the GT theory.

Holographic dual of the ${\cal N}=6$ mABJM theory is proposed in Ref.~\cite{Cheon:2011gv}, which is the $\mathbb{Z}_k $-quotient of the Lin-Lunin-Maldacena (LLM) geometry~\cite{Lin:2004nb} (see also  Ref.~\cite{Bena:2004jw}).
The proposal of the dual gravity gets much insights from the structure of the vacuum
space of the gauge theory. The dual gravity theories are not yet understood for the partially supersymmetric mABJM
theories. The ${\cal N}=4$ mABJM theory does not contain any Higgs vacuum
solution and does not seem to have a dual gravity theory related to
the LLM geometry. On the other hand, after the MP Higgsing and the dimensional uplift,
the resulting ${\cal N}=2^*$ mSYM theory turns out to be dual to the Pilch-Warner geometry in type IIB supergravity~\cite{Pilch:2000ue}.
It is interesting to figure out the
M-theory uplifting of this geometry and to identify the dual
geometry of the ${\cal N}=4$ mABJM theory. The ${\cal N}=2$ mABJM
theory has Higgs vacuum solutions but it is still unclear how to
modify the LLM geometry to obtain the corresponding dual gravity.

\section*{Acknowledgements}
This work was supported by the Korea Research Foundation Grant
funded by the Korean Government  with grant numbers 2011-0011660 (Y.K.), 2011-0009972 (O.K.), and 2009-0077423 (D.D.T.).

\appendix

\section{Four-dimensional Mass-deformed SYM Theories}\label{N=1star}
The four-dimensional ${\cal N}=1^*$ theory by Polchinski and Strassler~\cite{Polchinski:2000uf} is constructed by introducing a
mass-deformation to the ${\cal N}=4$ SYM theory. The action for the
latter is given by
\begin{align}\label{4dact}
{\cal \tilde L}=&{\rm tr}\Big[-\frac12\,F_{\alpha\beta}F^{\alpha\beta}-\tilde D^\alpha\tilde\Phi_a\tilde
D_\alpha\tilde\Phi_a+\frac{\tilde g^2}2\,[\tilde\Phi_a,\tilde\Phi_b]^2\nonumber\\
&~~~~+ \big(i\bar\psi_p\tilde\gamma^\alpha \tilde
D_\alpha\psi_p\big)-{\tilde g}\Big(\bar\psi_p\big(\Delta_a^{pq}\frac{1+\gamma_5}2+\bar\Delta_a^{pq}\frac{1-\gamma_5}2
\big)[\tilde\Phi_a,\psi_q]\Big)\Big],
\end{align}
where $\alpha,\beta=0,...,3$, $a,b=1,...,6$, $p,q=1,...,4$, $\psi_p$'s
 are Majorana fermions and $\tilde\Phi_a$'s are Hermitian scalar fields. $\Delta_a^{pq}=g_p\Delta_ag_q$ and
  $\bar\Delta_a^{pq}=g^\ast_p\Delta_ag^\ast_q$ are constants. $\Delta_a$
  are the gamma matrices of the six-dimensional Euclidean
  space, and $g_p$, $g_p^\ast$ are the eigenvectors of
  $\Gamma_\ast=-i\Delta_1...\Delta_6$ with eigenvalues +1, -1,
  respectively.  The covariant derivative is given by
 $\tilde D_\alpha=\partial_\alpha+i\tilde g[A_\alpha,.]$.
  The Clifford algebra for the gamma matrices is given by:
 $\{\tilde\gamma_\alpha,\tilde\gamma_\beta\}=-2\eta_{\alpha\beta}$ with the signature $\eta_{\alpha\beta}={\rm diag}(-1,1,1,...)$.

 The ${\cal N}=4$ supersymmetry transformation rules are
 \begin{align}\label{N=4susy}
 &\delta_\epsilon
 A_{\alpha}=i\bar\epsilon_p\tilde\gamma_\alpha\psi_p,
 \nonumber\\
 &\delta_\epsilon
 \tilde\Phi_a=i\bar\epsilon_p\big(\Delta_a^{pq}\frac{1+\gamma_5}2+\bar\Delta_a^{pq}\frac{1-\gamma_5}2
\big)\psi_q,
\\
&\delta_\epsilon\psi_p=iF_{\alpha\beta}\Sigma^{\alpha\beta}\epsilon_p+
\tilde \gamma^{\alpha}\tilde D_\alpha\tilde\Phi_a\big(\Delta_a^{pq}\frac{1+\gamma_5}2
+\bar\Delta_a^{pq}\frac{1-\gamma_5}2\big)\epsilon_q-{\tilde g}[\tilde\Phi_a,
\tilde\Phi_b]\big(\Delta^{pq}_{ab}\frac{1+\gamma_5}2
+\bar\Delta^{pq}_{ab}\frac{1-\gamma_5}2\big)\epsilon_q, \nn
 \end{align}
where the supersymmetry parameters $\epsilon_p$'s are Majorana fermions, and $\Delta_{ab}^{pq}=g^\ast_p\Sigma^{ab}g_q$,
$\bar\Delta^{ab}_{pq}=g_p\Sigma^{ab}g^\ast_q$ with $\Sigma^{ab} = -\frac{i}{4}[\Delta^a,\,\Delta^b]$.
After some algebra we obtain
 \begin{align}
 &\Delta^{pq}_{ab}=\frac{i}{4}\big(\bar\Delta_a^{po}\Delta_b^{oq}
 -\bar\Delta_b^{po}\Delta_a^{oq}\big),\nonumber\\
 &\bar\Delta^{pq}_{ab}=\frac{i}{4}\big(\Delta_a^{po}\bar\Delta_b^{oq}
 -\Delta_b^{po}\bar\Delta_a^{oq}\big),
 \end{align}
where the components of $\Delta_a$'s are given by
\begin{align}
&\Delta_1^{pq}-\bar\Delta_1^{pq}=2i\big(\delta_{p1}\delta_{q4}
-\delta_{p4}\delta_{q1}+\delta_{p2}\delta_{q3}
-\delta_{p3}\delta_{q2}\big),\quad\quad \Delta_1^{pq}+\bar\Delta_1^{pq}=0,\nonumber\\
&\Delta_2^{pq}-\bar\Delta_2^{pq}=2i\big(\delta_{p1}\delta_{q2}
-\delta_{p2}\delta_{q1}+\delta_{p3}\delta_{q4}
-\delta_{p4}\delta_{q3}\big),\quad\quad \Delta_2^{pq}+\bar\Delta_2^{pq}=0,\nonumber\\
&\Delta_3^{pq}-\bar\Delta_3^{pq}=2i\big(\delta_{p1}\delta_{q3}
-\delta_{p3}\delta_{q1}-\delta_{p2}\delta_{q4}
+\delta_{p4}\delta_{q2}\big),\quad\quad \Delta_3^{pq}+\bar\Delta_3^{pq}=0,\nonumber\\
&\Delta_4^{pq}+\bar\Delta_4^{pq}=-2\big(\delta_{p1}\delta_{q4}
-\delta_{p4}\delta_{q1}-\delta_{p2}\delta_{q3}
+\delta_{p3}\delta_{q2}\big),\quad\quad \Delta_4^{pq}-\bar\Delta_4^{pq}=0,\nonumber\\
&\Delta_5^{pq}+\bar\Delta_5^{pq}=2\big(\delta_{p1}\delta_{q2}
-\delta_{p2}\delta_{q1}-\delta_{p3}\delta_{q4}
+\delta_{p4}\delta_{q3}\big),\quad\quad~~ \Delta_5^{pq}-\bar\Delta_5^{pq}=0,\nonumber\\
&\Delta_6^{pq}+\bar\Delta_6^{pq}=-2\big(\delta_{p1}\delta_{q3}
-\delta_{p3}\delta_{q1}+\delta_{p2}\delta_{q4}
-\delta_{p4}\delta_{q2}\big),\quad\quad \Delta_6^{pq}-\bar\Delta_6^{pq}=0.
\end{align}
In the ${\cal N}=1^*$ theory,
without loss of generality we can choose $\epsilon=\epsilon_4$ as the
unbroken supersymmetry parameter with the other supersymmetry parameters set to zero. Then the supersymmetry
transformation rules in \eqref{N=4susy} are reduced to
\begin{align}\label{N=1*susy}
 &\delta_\epsilon
 A_{\alpha}=i\bar\epsilon\tilde\gamma_\alpha\lambda,\nonumber\\
 &\delta_\epsilon
 \tilde\Phi_a=i\bar\epsilon\big(\Delta_a^{4t}\frac{1+\gamma_5}2+\bar\Delta_a^{4t}\frac{1-\gamma_5}2
\big)\psi_t,\nonumber\\
 &\delta_\epsilon\psi_t=
 \tilde \gamma^{\alpha}\tilde D_\alpha\tilde\Phi_a\big(\Delta_a^{t4}\frac{1+\gamma_5}2
 +\bar\Delta_a^{t4}\frac{1-\gamma_5}2\big)\epsilon-{\tilde g}[\tilde\Phi_a,
 \tilde\Phi_b]\big(\Delta^{t4}_{ab}\frac{1+\gamma_5}2
 +\bar\Delta^{t4}_{ab}\frac{1-\gamma_5}2\big)\epsilon,\nonumber\\
 &\delta_\epsilon\lambda=iF_{\alpha\beta}\Sigma^{\alpha\beta}\epsilon-{\tilde g}[\tilde\Phi_a,
 \tilde\Phi_b]\big(\Delta^{44}_{ab}\frac{1+\gamma_5}2
 +\bar\Delta^{44}_{ab}\frac{1-\gamma_5}2\big)\epsilon,
 \end{align}
where $\lambda=\psi_4$ and $t=1,2,3$.

The mass-deformation
preserving the ${\cal N}=1$ supersymmetry is
\begin{align}\label{4dmdact}
{\cal L}_\mu={\rm tr}\big(-i\mu_{pq}\bar\psi_p\psi_q-M_{ab}\tilde\Phi_a\tilde\Phi_b
+i\tilde gT_{abc}\tilde\Phi_a[\tilde\Phi_b,\,\tilde\Phi_c]\big),
\end{align}
where
\begin{align}
\mu_{pq}={\rm diag}(\mu_1,\mu_2,\mu_3,0),\quad M_{ab}={\rm
diag}(\mu_1^2,\mu_3^2,\mu_2^2,\mu_1^2,\mu_3^2,\mu_2^2)
\end{align}
and the nonvanishing components of $T_{abc}$ are
\begin{align}
&T_{234}=\frac13(\mu_1-\mu_2-\mu_3),\quad
T_{126}=\frac13(\mu_1-\mu_2+\mu_3),\nonumber\\
&T_{135}=\frac13(\mu_1+\mu_2-\mu_3),\quad T_{456}=\frac13(\mu_1+\mu_2+\mu_3).
\end{align}
The modification to the fermionic variation is
\begin{align}\label{delta'}
\delta'_\epsilon\psi_p=\mu_{pq}\big(\Delta_a^{q4}\frac{1+\gamma_5}2
 +\bar\Delta_a^{q4}\frac{1-\gamma_5}2\big)\epsilon\tilde\Phi_a.
\end{align}
When $\mu_1 = \mu_2$ and $\mu_3=0$, we easily notice that the supersymmetry is
enhanced to ${\cal N}=2$. This gives the ${\cal N}=2^*$ theory discussed in Ref~\cite{Polchinski:2000uf}.

\subsection{Reduction to three dimensions}

In order to reduce the ${\cal N}=1^*$ theory to three dimensions we assume that the fields do not depend on the compactified direction. For the bosonic part, by introducing $V^{\frac12}A_{\alpha}=(A_\mu,~\phi)$ with $V$ the volume of
 the compactfied direction and $\mu,\nu=0,1,2$, we obtain
 \begin{align}
 VF_{\alpha\beta}F^{\alpha\beta}=F_{\mu\nu}F^{\mu\nu}+2F_{3\mu}F^{3\mu}=
 F_{\mu\nu}F^{\mu\nu}+2D^\mu\phi D_\mu\phi,
 \end{align}
and, by setting $V^{\frac12}\tilde\Phi_a=\Phi_a$ and
 $V^{-\frac12}\tilde g=g$, we have
 \begin{align}
 &V\tilde D^\alpha\tilde\Phi_a\tilde D_\alpha\tilde\Phi_a
 =D^\mu\Phi_aD_\mu\Phi_a-g^2[\phi,\Phi_a]^2,
 \quad V M_{ab}\tilde\Phi_a\tilde\Phi_b=M_{ab}\Phi_a\Phi_b
,\nonumber\\
&iV\tilde gT_{abc}\tilde\Phi_a[\tilde\Phi_b,\,\tilde\Phi_c]=
igT_{abc}\Phi_a[\Phi_b,\,\Phi_c],\quad V\tilde
g^2[\tilde\Phi_a,\tilde\Phi_b]^2=g^2[\Phi_a,\Phi_b]^2,
 \end{align}
  where the covariant derivative is given by
 $D_\mu=\partial_\mu-ig[A_\mu,.]$.
Using the relation $\int d^4x{\cal\tilde L}_{\rm bos}=\int d^3x{\cal L}_{\rm bos}$ and substituting the obtained results into the bosonic part of the four-dimensional action in \eqref{4dact} and \eqref{4dmdact}, we write the bosonic part of the Lagrangian density in three dimension as
\begin{align}\label{redN=1}
 {\cal L}_{\rm bos}=~&{\rm tr}\Big[-\frac12 F_{\mu\nu}F^{\mu\nu}-D^\mu\phi
 D_\mu\phi- D^\mu\Phi_aD_\mu\Phi_a+\frac12g^2\big([\phi,\Phi_a]^2+[\Phi_a,\phi]^2\big)\nonumber\\
 & +\frac12g^2[\Phi_a,\Phi_b]^2-M_{ab}\Phi_a\Phi_b+igT_{abc}\Phi_a[\Phi_b,\,\Phi_c]\Big]
 \nonumber\\
 =-&{\rm tr}\Big[\frac12 F_{\mu\nu}F^{\mu\nu}+ D^\mu\tilde X^i
 D_\mu\tilde X^i-\frac12g^2[\tilde X^i,\tilde X^j]^2+M_{ij}\tilde X^i\tilde X^j
 -igT_{ijk}\tilde X^i[\tilde X^j,\,\tilde X^k]\Big],
\end{align}
where $\tilde X^i=(\Phi_a,\phi)$ for $i=1,...,7$ are the seven
 transverse scalar fields and $M_{7i}=T_{7ij}=0$.

 For the fermionic part, we split the four-dimensional gamma matrices
 as follows
 \begin{align}\label{3dgamma}
 \tilde\gamma^0=\sigma_3\otimes i\sigma_2,\quad \tilde\gamma^1
 =\sigma_3\otimes
 \sigma_1,\quad\tilde\gamma^2=\sigma_3\otimes \sigma_3,\quad
 \tilde\gamma^3=\sigma_1\otimes \mathbb{I},
 \end{align}
 whereas the three-dimensional gamma matrices are given by
 \begin{align}
 \gamma^0=i\sigma_2,\quad \gamma^1
 = \sigma_1,\quad\gamma^2=\sigma_3.
 \end{align}
 The four-dimensional Majorana spinor has the expansion
\begin{align}
 V^{\frac12}\psi_p=\sum_{r=1}^2e^r\otimes\psi_p^r,
 \end{align}
 where $e^r$'s form the basis of $\mathbb{R}^2$ and $\psi_p^r$'s are Majorana
 spinors in three dimensions. With $\gamma_5=-i\tilde\gamma_0\tilde\gamma_1\tilde\gamma_2\tilde\gamma_3$, their chiral components are written in terms of the three-dimensional Majorana spinors as
\begin{align}
&V^{\frac12}\psi_p^{+}=V^{\frac12}\frac{1+
\gamma_5}2\psi_p=\frac{1-\sigma_2
\otimes\mathbb{I}}2e^r\otimes\psi_p^r,\nonumber\\
&V^{\frac12}\psi_p^{-}=V^{\frac12}\frac{1-
\gamma_5}2\psi_p=\frac{1+\sigma_2
\otimes\mathbb{I}}2e^r\otimes\psi_p^r.
\end{align}
Finally, the covariant derivatives are given by
\begin{align}
 &V^{\frac12}\tilde D_\mu\psi_p=e^r\otimes D_\mu\psi_p^r,\quad\quad
 V^{\frac12}\tilde D_3\psi_p=ige^r\otimes[\phi, \psi_p^r],
 \end{align}
and the Dirac conjugation becomes
\begin{align}\label{diracconj}
V^{\frac12}\bar\psi_p=V^{\frac12}\psi_p^\dagger\tilde\gamma^0=
(e^r\otimes\psi_p^r)^\dagger \sigma_3\otimes i\sigma_2.
\end{align}

Similar to the bosonic part, we substitute the results \eqref{3dgamma}--\eqref{diracconj} into every term of  the fermionic part in \eqref{4dact} and \eqref{4dmdact} and use $\int d^4x{\cal\tilde L}_{\rm ferm}=\int d^3x{\cal L}_{\rm ferm}$. Computation of the kinetic term leads to
\begin{align}\label{3dkin}
&V\bar\psi_p\tilde\gamma^\alpha \tilde
D_\alpha\psi_p=V\bar\psi_p\tilde\gamma^\mu \tilde
D_\mu\psi_p+V\bar\psi_p\tilde\gamma^3 \tilde D_3\psi_p\nonumber\\
&=(e^{r\dagger}\otimes\psi_p^{r\dagger})\Big((\sigma_3\otimes\gamma^0)
(\sigma_3\otimes\gamma^\mu) e^s\otimes
D_\mu\psi_p^s+ig(\sigma_3\otimes\gamma^0)
(\sigma_1\otimes\mathbb{I}) e^s\otimes
[\phi,\psi_p^s]\Big)\nonumber\\
&=\bar\psi_p^r\gamma^\mu D_\mu\psi_p^r
+ig\gamma^0_{rs}\bar\psi_p^r[\phi,\psi_p^s],
\end{align}
where $\bar\psi_p^r=\psi_p^{r\dagger} \gamma^0$ is the Dirac
conjugation in three dimensions, $\gamma^0_{rs}=e^{r\dagger}\gamma^0
e^s=ie^{r\dagger}\sigma_2 e^s$, and we have used
$e^{r\dagger}e^s=\delta^{rs}$.
 The fermionic mass term and the Yukawa-type interaction terms are
\begin{align}\label{3dpot}
&iV\mu_{pq}\bar\psi_p\psi_q=i\mu_{pq}\psi_p^1\gamma^0\psi_q^1-i\mu_{pq}\psi_p^2\gamma^0
\psi_q^2,
\nn \\
&V\tilde g\Delta_a^{pq}\bar\psi_p[\tilde\Phi_a,\psi_q^{+}]= \frac
g2\Delta_a^{pq}\big(i\gamma^1_{rs}+\gamma^2_{rs}\big)\bar\psi_p^r[\Phi_a,\psi_q^{s}],
\nn \\
&V\tilde g\bar\Delta_a^{pq}\bar\psi_p[\tilde\Phi_a,\psi_q^{-}]= \frac
g2\bar\Delta_a^{pq}\big(-i\gamma^1_{rs}+\gamma^2_{rs}\big)\bar\psi_p^r[\Phi_a,\psi_q^{s}],
\end{align}
where $\gamma^1_{rs}=e^{r\dagger}\gamma^1 e^s=e^{r\dagger}\sigma_1
e^s$ and $\gamma^2_{rs}=e^{r\dagger}\gamma^2
e^s=e^{r\dagger}\sigma_3 e^s$.

 Collecting all the terms in \eqref{3dkin}--\eqref{3dpot} we have
\begin{align}
{\cal L}_{\rm ferm}=~~& {\rm tr}\Big[i\bar\psi_p^r\gamma^\mu
D_\mu\psi_p^r-g\gamma^0_{rs}\bar\psi_p^r[\phi,\psi_p^s]-i\mu_{pq}\psi_p^1\gamma^0\psi_q^1+i\mu_{pq}\psi_p^2\gamma^0\psi_q^2\nonumber\\
&-\frac
g2\Big[i\gamma^1_{rs}\big(\Delta_a^{pq}-\bar\Delta_a^{pq}\big)
+\gamma^2_{rs}\big(\Delta_a^{pq}+\bar\Delta_a^{pq}\big)\Big]\bar\psi_p^r
[\Phi_a,\psi_q^{s}]\Big].
\end{align}
This is rewritten as
\begin{align}\label{redN=1fer}
{\cal L}_{\rm ferm}={\rm tr}\Big[i\bar\psi_p^r\gamma^\mu
D_\mu\psi_p^r-i\mu_{pq}\psi_p^1\gamma^0\psi_q^1+i\mu_{pq}\psi_p^2\gamma^0\psi_q^2-g(M_i)_{rs}^{pq}\bar\psi_p^r[\tilde
X^i,\psi_q^s]\Big],
\end{align}
by using seven-dimensional index $i=1,...,7$ and the corresponding Clifford algebra of the Euclidean Gamma matrices
\begin{align}
\{M_i,M_j\}^{pq}_{rs}=-2\delta_{ij}\delta_{rs}\delta^{pq},
\end{align}
defined by
\begin{align}
&(M_7)_{rs}^{pq}=\gamma^0_{rs}\delta^{pq},\nonumber\\
& (M_i)_{rs}^{pq}=
i\gamma^1_{rs}\Delta_i^{pq}=-i\gamma^1_{rs}\bar\Delta_i^{pq},\quad
{\rm
for}~~ i=1,2,3,\nonumber\\
& (M_i)_{rs}^{pq}=
\gamma^2_{rs}\Delta_i^{pq}=\gamma^2_{rs}\bar\Delta_i^{pq},\quad {\rm
for}~~ i=4,5,6.
\end{align}

The Lagrangians in \eqref{redN=1} and \eqref{redN=1fer} are identical to the
three-dimensional ${\cal N}=2$ mSYM theory with the mass parameters adjusted to the `case II' of \eqref{class}. Similarly, the dimensional reduction
of the ${\cal N}=2^*$ theory gives the ${\cal N}=4$ mSYM theory discussed in
subsection 4.1.

\subsection{Reduction of the supersymmetry variations}
Let us subsequently discuss the supersymmetry variation after dimensional reduction in this subsection. The supersymmetry parameters in \eqref{N=4susy} and their chiral components have an expansion in three-dimensions,
 \begin{align}
\epsilon_p=e^r\otimes\epsilon_p^r~,\quad\quad\epsilon_p^{+}=\frac{1-\sigma_2
\otimes\mathbb{I}}2e^r\otimes\epsilon_p^r~,\quad\quad\epsilon_p^{-}=\frac{1+\sigma_2
\otimes\mathbb{I}}2e^r\otimes\epsilon_p^r~,
\end{align}
where $\epsilon_p^r$'s are three-dimensional Majorana spinors. Since the case of our consideration sets only $\epsilon_4$ nonvanishing, the supersymmetry
variation of the bosonic and fermionic fields in \eqref{N=1*susy} becomes
 \begin{align}\label{3dN=1susy}
 &\delta_\epsilon
 \phi=i\gamma^0_{rs}\bar\epsilon^r\lambda^s,\nonumber\\
 &\delta_\epsilon
 A_{\mu}=i\bar\epsilon^r\gamma_\mu\lambda^r,\nonumber\\
 &\delta_\epsilon
 \Phi_a=\frac i2\Big(i\gamma^1_{rs}\big(\Delta_a^{4t}-\bar\Delta_a^{4t}\big)
+\gamma^2_{rs}\big(\Delta_a^{4t}+\bar\Delta_a^{4t}\big)\Big)\bar\epsilon^r
\psi_t^{s},\nonumber\\
 &\delta_\epsilon\psi_t^r=\frac12\Big(i\gamma^1_{rs}(\Delta_a^{t4}-\bar\Delta_a^{t4})+\gamma^2_{rs}
(\Delta_a^{t4}+\bar\Delta_a^{t4})\Big)D_\mu\Phi_a\gamma^\mu
\epsilon^{s}\nonumber\\
  &~~~~~~~~~-\frac g4\Big(i\gamma^1_{rs}(\Delta_a^{t4}+\bar\Delta_a^{t4})+
  \gamma^2_{rs}(\Delta_a^{t4}-\bar\Delta_a^{t4})\Big)\big([\Phi_a,\phi]-
  [\phi,\Phi_a]\big)\epsilon^{s}
 \nonumber\\
   &~~~~~~~~-\frac g2[\Phi_a,\Phi_b]\Big((\Delta^{t4}_{ab}+
   \bar\Delta^{t4}_{ab})\epsilon^{r}+i\gamma^0_{rs}
   (\Delta^{t4}_{ab}-\bar\Delta^{t4}_{ab})\epsilon^{s}\Big),\nonumber\\
   &\delta_\epsilon\lambda^r=iF_{\mu\nu}\sigma^{\mu\nu}\epsilon^r
 +\gamma^0_{rs}D_\mu\phi\gamma^\mu\epsilon^s-\frac g2[\Phi_a,\Phi_b]\Big((\Delta^{44}_{ab}+
   \bar\Delta^{44}_{ab})\epsilon^{r}+i\gamma^0_{rs}
   (\Delta^{44}_{ab}-\bar\Delta^{44}_{ab})\epsilon^{s}\Big),
  \end{align}
where $\sigma^{\mu\nu}=\frac1{4i}(\gamma^\mu\gamma^\nu-\gamma^\nu\gamma^\mu)$.
By using the compact notation introduced in the pervious subsection \eqref{3dN=1susy} is simplified as
\begin{align}
  &\delta_\epsilon
 A_{\mu}=i\bar\epsilon^r\gamma_\mu\lambda^r,
\nonumber\\
 &\delta_\epsilon
 \tilde X^i=i(M_i)^{4q}_{rs}\bar\epsilon^r
\psi_q^{s},
\nonumber\\
 &\delta_\epsilon\psi_t^r= (M_i)^{t4}_{rs}D_\mu\tilde X^i\gamma^\mu\epsilon^s
 -g(M_{ij})_{rs}^{t4}[\tilde X^i,\tilde X^j]\epsilon^{s},
\nonumber\\
 &\delta_\epsilon\lambda^r=iF_{\mu\nu}\sigma^{\mu\nu}\epsilon^r
  -g(M_{ij})_{rs}^{44}[\tilde X^i,\tilde X^j]\epsilon^{s},
  \end{align}
where $(M_{ij})_{rs}^{pq}= \frac i4\Big(M_iM_j- M_jM_i\Big)^{pq}_{rs}.$
Similar computation to the fermionic mass part \eqref{delta'} also provides simpler expression,
\begin{align}
\delta'_\epsilon\psi_p^r=\mu_{pq}\big(\Delta_a^{q4}\frac{1+\gamma_5}2
 +\bar\Delta_a^{q4}\frac{1-\gamma_5}2\big)\epsilon\tilde\Phi_a
 =\mu_{pq}(M_i)^{q4}_{rs}\tilde X^i\epsilon^s.
\end{align}

\end{document}